\documentclass[graybox]{svmult}

\usepackage{url}       
\usepackage{mathptmx}       
\usepackage{helvet}         
\usepackage{courier}        
\usepackage{type1cm}        
\usepackage{makeidx}         
\usepackage{graphicx}        
\usepackage{multicol}        
\usepackage[bottom]{footmisc}

\makeindex             

\begin{document}

\title*{Charge Kondo Effect in Thermoelectric Properties of Lead Telluride doped with Thallium Impurities}
\titlerunning{Charge Kondo Effect in PbTe doped with Tl Impurities}
\author{T. A. Costi and V. Zlati\'c}
\institute{T. A. Costi \at 
Peter Gr\"{u}nberg Institut and Institute for Advanced Simulation, 
Research Centre J\"{u}lich, 52425 J\"ulich, Germany,
\email{t.costi@fz-juelich.de}
\and V. Zlati\'c
\at Institute of Physics, Zagreb POB 304, Croatia and J. Stefan Institute, SI-1000 Ljubljana, Slovenia 
\email{zlatic@ifs.hr}}

\maketitle

\abstract{
We investigate the thermoelectric properties of PbTe doped with a small
concentration $x$ of Tl impurities acting as acceptors and described by
Anderson impurities with negative on-site (effective) interaction. The resulting
charge Kondo effect naturally accounts for a number of the low temperature
anomalies in this system, including the unusual doping dependence of the carrier
concentration, the Fermi level pinning and the self-compensation effect. The
Kondo anomalies in the low temperature resistivity at temperatures $T\leq 10\, {\rm K}$
and the $x$-dependence of the residual resistivity are also in good agreement with 
experiment. Our model also captures the qualitative aspects of the 
thermopower at higher temperatures $T>300\, {\rm K}$ for high dopings
($x>0.6\%$) where transport is expected to be largely dominated by carriers in the 
heavy hole band of PbTe.
}

\section{Introduction}
\label{sec:1}
The ever growing demand for energy has increased interest in 
materials with high thermoelectric (TE) efficiency. 
A remarkable feature of TE devices is that they directly 
convert heat into electrical energy or use 
electricity to pump heat by circulating the electron fluid between a hot and 
a cold reservoir \cite{callen.85,ioffe.54}.
The absence of mechanical parts makes TE devices very reliable 
but their low efficiency restricts 
applications to rather specialized fields. Today, they are mainly 
used in situations where 
reliability is the most important factor, e.g., for electricity 
generation in remote regions 
or for vibrationless cooling of sensitive devices. 

The efficiency of TE materials is related to their figure-of-merit, 
$ZT=P\; T/\kappa $,  where $P$ is the power factor and $\kappa $ is the 
thermal conductivity. Current applications are based on 
semiconducting materials that have a state-of-the-art $ZT$ 
of around one \cite{snyder.08, kanatzidis.10}, 
which is not sufficient for wide-spread use.  The production of cheap 
TE materials with a room  temperature value of $ZT\geq 2$  
would be a major technological breakthrough. 

The efforts to increase $ZT$ follow two general strategies. 
The one tries to reduce $\kappa$ and increase the efficiency by 
minimizing the thermal losses through the device. The other tries to 
increase $P=S^2 \sigma$, were  $S(T)$ is the Seebeck coefficient and 
$\sigma(T)$ the electrical conductivity. Recent advances that achieved
$ZT\geq 2$ are mostly due to the reduction of $\kappa$ by nanostructuring, 
i.e., by using 
multilayered materials (for review see \cite{kanatzidis.10}) and 
nanocomposites \cite{minnich.08}.
The enhancement of $ZT$ via the power factor is less straightforward, 
because an increase of $S(T)$ is usually accompanied by a reduction 
of $\sigma$ and vice versa \cite{mahan.97}. The reason is that $\sigma(T)$ 
is mainly determined by the value of the density of states $\cal N(\omega)$ 
and the transport relaxation time $\tau(\omega)$ at  the chemical potential $\mu$, 
while $S(T)$ depends sensitively on the symmetry of $\cal N(\omega)$ 
and $\tau(\omega)$ around $\mu$.
Thus, one way to a larger power 
factor is by increasing the asymmetry 
of $\cal N(\omega)$ in the proximity of $\mu$ by, say, 
confining the electron gas to low dimensions \cite{hicks.93} 
or by introducing 
resonant states within the Fermi window by doping \cite{mahan_sofo.96}. 
This strategy could work well for 
materials in which the charge carriers behave 
as non-interacting Fermions and the energy dependence of $\tau(\omega)$ 
is negligibly small \cite{singh.10}.

The other route to large $P$  is via electron correlation which 
makes $\tau(\omega)$ strongly energy dependent, as 
in heavy fermions \cite{mahan.97,zlatic.05}, e.g., YbAl$_3$, 
transition metal oxides, e.g. Na$_x$CoO$_2$ \cite{terasaki.97,wisgott.10},
or in correlated semiconductors, such as FeSi \cite{bentien.06,sales.94}. 
However, these materials  have large $\kappa$  such that $ZT$ is too small 
for applications, except, perhaps, at very low temperatures. 
At high temperatures, the correlation effects are usually less pronounced 
and the appropriate materials are not easy to find. 
Considering the complexity of the problem, it is clear that the search for better 
thermoelectrics could benefit considerably  from an understanding 
of basic microscopic mechanisms that govern 
the thermoelectric response of a given material.

One of the most interesting thermoelectric materials is Pb$_{1-x}$Tl$_{x}$Te, 
where $x$ is the concentration of Tl impurities 
\cite{ravich.98,volkov.02,heremans.08}.
The parent compound PbTe is a narrow gap semiconductor with a gap of $190\,{\rm meV}$
and is itself a good thermoelectric material with a thermopower of around 
$300\mu {\rm V/K}$ at $800\, {\rm K}$. Upon doping with Tl impurities, which act as acceptors 
\cite{ravich.98,volkov.02}, a large thermopower persists at high temperature, while
a number of anomalies appear in the low temperature properties. Remarkable, these 
depend sensitively on the Tl concentration with a qualitatively 
different behavior below and above a critical concentration $x_c\simeq 0.3$  at\% Tl. 
For example, Pb$_{1-x}$Tl$_{x}$Te becomes superconducting for $x>x_{c}$ 
with a transition temperature T$_c(x)$ increasing linearly with $x$ and 
reaching $1.5\,{\rm K}$ at $x=1.5$ at\% Tl \cite{chernik.81,matsushita.06}. 
This is surprisingly high given the low 
carrier density of 10$^{20}$ holes/cm$^3$. Measurements of the Hall number $p_H=1/R_H e$ 
\cite{matsushita.06} indicate that the number of holes grows linearly with 
$x$ for $x \leq x_c$, whereas for $x>x_{c}$ the number of holes remains almost 
constant: the system exhibits ``self-compensation'' and the chemical 
potential is pinned to a value $\mu\simeq \mu^{*}=220\,{\rm meV}$
\cite{matsushita.06,murakami.96,kaidanov.89}. 
Transport measurements show anomalous behavior at low temperatures: 
while for $x<x_{c}$, the residual resistance $\rho_0$ is very small 
and almost constant \cite{matusiak.09}, for $x>x_{c}$, 
$\rho_{0}$ increases approximately linearly with $x$. For $x<x_{c}$, the
resistivity, $\rho(T)$ exhibits a positive slope at low temperature
\cite{matsushita.05}, 
while for $x>x_{c}$, the slope is negative and a Kondo like impurity 
contribution $\rho_{\rm imp}(T)$ is observed for $T\leq 10\, {\rm K}$. 
The origin of this anomaly is not due to magnetic impurities, since
the susceptibility is diamagnetic \cite{matsushita.05}. Finally, 
at low Tl concentrations, $S(T)$ shows a sign-change before growing 
to large positive values at room temperatures \cite{matusiak.09}.

Several models have been proposed to explain the anomalous properties
of Pb$_{1-x}$Tl$_{x}$Te. The simplest assumes that doping with Tl gives rise
to non-interacting resonant levels close to the top of the p-valence band of 
PbTe \cite{ravich.98}. Density functional theory (DFT) calculations   
confirm the presence of such states with Tl {\it s}-character \cite{xiong.10}. 
In the static mixed valence model \cite{drabkin.81,volkov.02}, 
Tl impurities, known to be valence skippers in compounds, are 
assumed to dissociate into energetically close Tl$^{1+}$ (6{\it s}$^2${\it p}$^1$) 
and Tl$^{3+}$ (6{\it s}$^0${\it p}$^3$) ions,
while the Tl$^{2+}$ (6{\it s}$^1${\it p}$^2$) configuration lies higher
in energy (consistent with DFT calculations \cite{weiser.79}). 
With a strong electron-phonon interaction such a model results 
in negative on-site $U$. This model provides
a natural explanation for the observed self-compensation and the diamagnetic
behavior of Pb$_{1-x}$Tl$_{x}$Te \cite{drabkin.81}. In addition, it provides
a mechanism for the onset of superconductivity. Dynamic fluctuations 
between the Tl$^{1+}$ and Tl$^{3+}$ valence 
states results in the negative $U$ Anderson model \cite{anderson.75} which 
supports a charge Kondo effect \cite{taraphder.91,andergassen.11} 
and has been proposed to explain the anomalous properties of Pb$_{1-x}$Tl$_{x}$Te 
\cite{matsushita.05,matsushita.06,matusiak.09,erickson.10,malshukov.91,dzero.05,costi.12}, 
including the superconductivity for $x>x_{c}$ \cite{malshukov.91,dzero.05}. 
A coupling of the Tl ions to the lattice has also been considered \cite{shelankov.87,martin.97}.
Recent ARPES data on a $0.5\%$ sample \cite{nakayama.08} 
are also consistent with a negative $U$ Anderson model.

As regards the thermoelectric properties of interest for applications, 
increasing the Tl concentration beyond 1 at\% gives rise to a 
room-temperature $ZT$ 
which is surprisingly large for a bulk material \cite{heremans.08}. 
The data show that the thermopower $S(T)$ increases almost linearly at low 
temperatures \cite{matusiak.09} and  saturates above 400 K 
at rather high values \cite{heremans.08}. 
The fact that the enhancement of $ZT$ occurs at high temperatures and 
that it is mainly due to an increased power factor makes this material 
particularly interesting not just from the practical but also
from the theoretical point of view.

In this paper, we use the numerical renormalization group approach (NRG) \cite{nrg}
to calculate the anomalous frequency and temperature dependent transport 
time, $\tau(\omega,T)$, of electrons scattering from Tl impurities 
described as negative-$U$ Anderson impurities. We show that this model
explains the unusual concentration and temperature dependent   
properties of Pb$_{1-x}$Tl$_{x}$Te. We take full account
of charge neutrality and include the realistic band structure of PbTe 
\cite{singh.10}. We find that both correlation and band effects are important
in accounting for the experimental data. We shall also show, that our 
single-band Anderson impurity model qualitatively reproduces the measured 
behaviour of the thermopower over a wide temperature range in the regime 
where transport is mainly due to carriers in a single heavy hole like 
sub-band of the PbTe valence band. At lower dopings, the population of 
two valence sub-bands of PbTe requires a further generalization of the 
model in order to describe the thermopower also in this regime (see 
discussion in Sec.~\ref{sec:9}). This, however, is beyond the scope of 
the present paper. 
\section{Negative-$U$ Anderson Model for Tl impurities in PbTe}
\label{sec:2}
We consider $n$ Tl impurities in a PbTe crystal with $N$ Pb sites
described by the Hamiltonian $H=H_{\rm band} +H_{\rm imp}+H_{\rm hyb}$, where 
\begin{eqnarray}
\label{H_band}
H_{\rm band} &=&  \sum_{{\bf k}\sigma} (\epsilon_{\bf k} -\mu) 
c^{\dagger}_{ {\bf k}\sigma}c_{ {\bf k}\sigma}~, \\
\label{H_imp}
H_{\rm imp}&=&  (\epsilon_0 -\mu) \sum_{i=1\sigma}^{n} 
\hat n_{is\sigma} + U \sum_{i=1}^{n}
n_{is\uparrow} n_{is\downarrow}~,\\
\label{H_hyb}
H_{\rm hyb} &=& \sum_{i=1}^{n}\sum_{k\sigma}V_{{\bf k}}
(c_{{\bf k}\sigma}^{\dagger}s_{i\sigma} + h.c.)~.
\end{eqnarray}
$c^{\dagger}_{{\bf k}\sigma} $ creates an electron
in the valence p-band at energy $\epsilon_{\bf k}$, 
 $\hat{n}_{is\sigma} =s^{\dagger}_{i\sigma} s^{}_{i\sigma}$ is the number operator for 
a Tl {\it s}-electron at site $i$ with spin $\sigma$ and energy $\epsilon_0$,  
$U$ is the (negative) on-site interaction, and $V_{{\bf k}}$ is the matrix element for
the {\it s}-{\it p} interaction. Its strength is characterized by the 
hybridization function 
$\Delta(\omega)=\pi \sum_{\bf k}|V_{{\bf k}}|\delta(\omega-\epsilon_{{\bf k}\sigma})$.
We neglect the ${\bf k}$-dependence of $V_{{\bf k}}$ setting $V_{{\bf k}}=V_{0}$ but
keep the full energy dependence of $\Delta(\omega)$ by using the p-band 
density of states ${\cal N}(\omega)=\sum_{\bf k}\delta(\omega-\epsilon_{\bf k})$ 
calculated
from DFT and including relativistic effects \cite{singh.10}.
The chemical potential $\mu$ determines $n_e=\frac{1}{N}\sum_{{\bf k}\sigma}
\langle c^{\dagger}_{ {\bf k}\sigma}c_{ {\bf k}\sigma}\rangle
$ and $n_s=\frac{1}{n}\sum_{i=1}^{n}\sum_{\sigma}\langle n_{is\sigma}\rangle$,  
the average number of {\it p} and {\it s} electrons per site. We denote by
$x=n/N$ the concentration of Tl impurities. Since Tl acts as an acceptor, 
the ground state 
corresponds to the Tl$^{1+}$ ($n_s=2$) configuration and the Tl$^{3+}$ ($n_s=0$) 
configuration is split-off from the ground state by the energy 
$\delta =E({\rm Tl}^{3+})-E({\rm Tl}^{1+})>0$. A concentration 
$x$ of Tl impurities accommodates $x(n_{s}-1)$ electrons (per Tl site), 
where the number of accepted 
electrons in the 6{\it s} level of Tl is measured relative to 
the neutral Tl$^{2+}$ ({\it s}$^{1}$) 
configuration having $n_{s}=1$. These electrons are removed from 
the valence band leaving
behind $n_{0}=1-n_{e}$ holes. 
Thus, the charge neutrality condition reads \cite{dzero.05}
\begin{equation} 
\label{eq:neutrality}
n_{0}=x (n_{s}-1)~,
\end{equation}
which for a given $x$ and temperature $T$ has to be satisfied by 
adjusting the hole chemical potential $\mu$. 
Here, we neglect inter-impurity interactions and solve $H$ 
for a finite number of independent negative-$U$ 
centres by using the NRG \cite{nrg}. For each $x$ and each $T$ we need to satisfy 
(\ref{eq:neutrality}) by self-consistently determining the chemical potential. 
In practice we found it more efficient to solve the negative-$U$ Anderson model 
on a dense grid of $256$ chemical potentials about $\mu=\mu^{*}$ 
and to subsequently convert these by interpolation to a fixed particle 
number (i.e. obeying Eq.~(\ref{eq:neutrality}) for each $T$ at given $x$).

\section{Transport coefficients}
\label{sec:3}

The transport coefficients for scattering from a dilute concentration $x$ of {\rm Tl} 
impurities are obtained from the Kubo formula \cite{mahan.90}. 
The electrical resistivity and the thermopower are defined by the usual expressions,
\begin{eqnarray}
\label{eq: conductivity}
{\rho_{\rm imp}}(T)&=&\frac{1}{e^2 L_{11}}~,\\
\label{eq: thermopower}
S(T)&=&-\frac{1}{|e|T}\frac{L_{12}}{L_{11}}~,
\end{eqnarray}
where $ L_{11}$ and $ L_{12}$ are given by the static limits of the
current-current and current-heat current correlation functions,
respectively. In the absence of non-resonant scattering the vertex corrections
vanish and the transport integrals can be written as \cite{mahan.90,costi.94}, 
\begin{equation}
\label{eq: lij_final}
L_{ij} = \sigma_{0}\int_{-\infty}^{\infty}d\omega
\left (
-\frac{df(\omega)}{d\omega} \right ){\cal N}(\omega)\tau(\omega,T)\omega^{i+j-2},
\end{equation}
where $\sigma_{0}$ is a material-specific constant,
$f(\omega)=1/[1+\exp(\omega/k_B T)]$ is the Fermi function,
$\tau(\omega,T)$ is the conduction-electron transport time\cite{costi.94}
\begin{equation}
\label{eq: tau}
\frac{1}{\tau(\omega,T)} = 2\pi c_{\rm imp}  V_{0}^2 A(\omega,T),
\end{equation}
and $A(\omega,T)=\mp \frac{1}{\pi}\mbox{Im}\; G(\omega \pm i 0^{+})$ is
the  spectral function of {\it s\/}-holes. The number of {\rm Tl} impurities 
$c_{\rm imp}$ per {\rm cm}$^3$ is related to $x$ in  \% by 
$c_{\rm imp}=1.48\times 10^{20}x$ (using the lattice constant 
$a_{0}=6.46\times 10^{-8}${\rm cm} of the PbTe rocksalt structure).
Eqs.(\ref{eq: lij_final})  and (\ref{eq: tau}) show that ${\rho_{\rm imp}}(T)$ depends 
strongly on the value of $A(\omega,T)$ around $\omega\simeq\mu$ 
and that  the sign and the magnitude of $S(T)$ follow from  the shape 
of ${\cal N}(\omega)A(\omega,T)$ within the Fermi window $|\omega|\leq 2k_BT$.
Since PbTe has an unusual non-parabolic band structure close to the top the 
valence band \cite{singh.10}, and since this region is accessible with Tl dopings $x\leq 1\%$, one
sees that the thermopower in particular will be sensitive to details of
both ${\cal N}(\omega)$ and $A(\omega,T)$.

\section{Choice of Model Parameters}
\label{sec:4}
Low temperature Hall effect and tunneling experiments for samples with $x>x_{c}$ 
provide an estimate for $\mu^{*}\approx 220\,{\rm meV}$ that we use.
Other parameters such as $\Delta_{0}=\Delta(\mu^{*})$, required to fix the hybridization
function $\Delta(\omega)$, and $U$ are unknown, or they depend strongly on the 
interpretation of experiments (see discussion below). 
However, the measured Kondo-like resistivity for $x\ge x_{c}$ at 
low temperatures requires that $|U|\gg \Delta_{0}$. 
We take $U/\Delta_{0}=-8$. As argued in \cite{costi.12}, it is possible
to estimate  $U$ and $\Delta_{0}$ by interpreting the point contact 
measurements \cite{murakami.96} within our model and by assuming that the
Kondo scale $T_{\rm K}$ in the resistivity anomaly at large dopings is comparable
to the $T_{c}\sim 1.5 {\rm K}$, resulting in $U=-30\, {\rm meV}$ 
and $\Delta=2.7\, {\rm meV}$. While, the precise values of these parameters
are required to consistently explain all measurements, in this paper we
shall choose different values to those in \cite{costi.12} in order to 
substantiate the claim made there, that the qualitative aspects of those results remain 
the same for $10\, {\rm meV} \leq |U| \leq 220\, {\rm meV}$ (and $U/\Delta_{0}=-8$).
Hence, we shall choose $\Delta_{0}=13.75\,{\rm meV}$ and $U=-110\,{\rm meV}$ resulting in 
$T_{\rm K}\approx 14\,{\rm K}$, where the Kondo scale $T_{\rm K}$ is defined via the
resistivity $\rho_{\rm imp}(T_{\rm K})=\rho_{\rm imp}(T=0)/2$ for $x\gg x_{c}$. This
is the same order of magnitude as the perturbative scale
$T_{\rm K}'=\Delta_{0} (U/2\Delta_{0})^{1/2}\exp(-\pi U/8\Delta_{0})=0.0864\Delta_{0}$ 
for the negative-$U$ Anderson model \cite{andergassen.11,hewson.97}.

\section{Qualitative considerations}
\label{sec:5}
Before presenting our results, it is instructive to consider the atomic limit $V_{0}=0$. 
For $x=0$ the chemical potential lies in the gap between the valence and conduction bands.
For finite but very small $x$ each Tl impurity accepts one electron, 
i.e. $n_{s}\approx 2$ and $n_{0}\approx x$ grows linearly with $x$. At the same time, the 
chemical potential shifts downwards into the valence band $\mu < E_{v}$, where $E_{v}$ denotes
the top of the valence band. This implies that the splitting $\delta(\mu)=-(2(\epsilon_0-\mu) + U)$ 
between donor and acceptor configurations {\em decreases}. Eventually, at a critical concentration
$x=x_{c}$, the chemical potential reaches $\mu=\mu^{*}=\varepsilon_{0}+U/2$ where 
$\delta(\mu)=2(\mu -\mu^{*})$ vanishes and the system is in a mixed valence state where 
the Tl$^{1+}$ and Tl$^{3+}$ configurations are degenerate. In this situation
$n_{s}=1$, and any further doping with Tl cannot increase the hole carrier density beyond
the value $n_{0}(\mu^{*})$, i.e. one has self-compensation with a pinning 
of the chemical potential to $\mu^{*}$ \cite{dzero.05}. Experiments show that the properties of  
Pb$_{1-x}$Tl$_{x}$Te change dramatically for $x>x_c\approx 0.3$ \% but a static mixed 
valence state implied by the above atomic limit cannot capture many of these, for example, 
the anomalous upturn of the resistivity at $T<10\, {\rm K}$.

For finite $V_{0}$, 
quantum fluctuations between the degenerate states Tl$^{1+}$ and Tl$^{3+}$ at
$\mu=\mu^{*}$ become important and lead to a charge Kondo effect\cite{taraphder.91}. 
This significantly affects
all static and dynamic properties \cite{andergassen.11} and needs to be taken into 
account in describing the experiments. This charge Kondo effect is important also for $\mu>\mu^{*}$, since 
a finite charge splitting $\delta(\mu)> 0 $ in the negative-$U$ Anderson model 
is similar to a Zeeman splitting in the conventional spin Kondo effect \cite{iche.72}. The latter is known
to drastically influence all properties. Thus, for the whole range of concentrations $x$, one
expects dynamic fluctuations to play an important role in the properties of Pb$_{1-x}$Tl$_{x}$Te.

\section{Carrier Concentration}
\label{sec:6}
\begin{figure}
\includegraphics[width=\linewidth,clip]{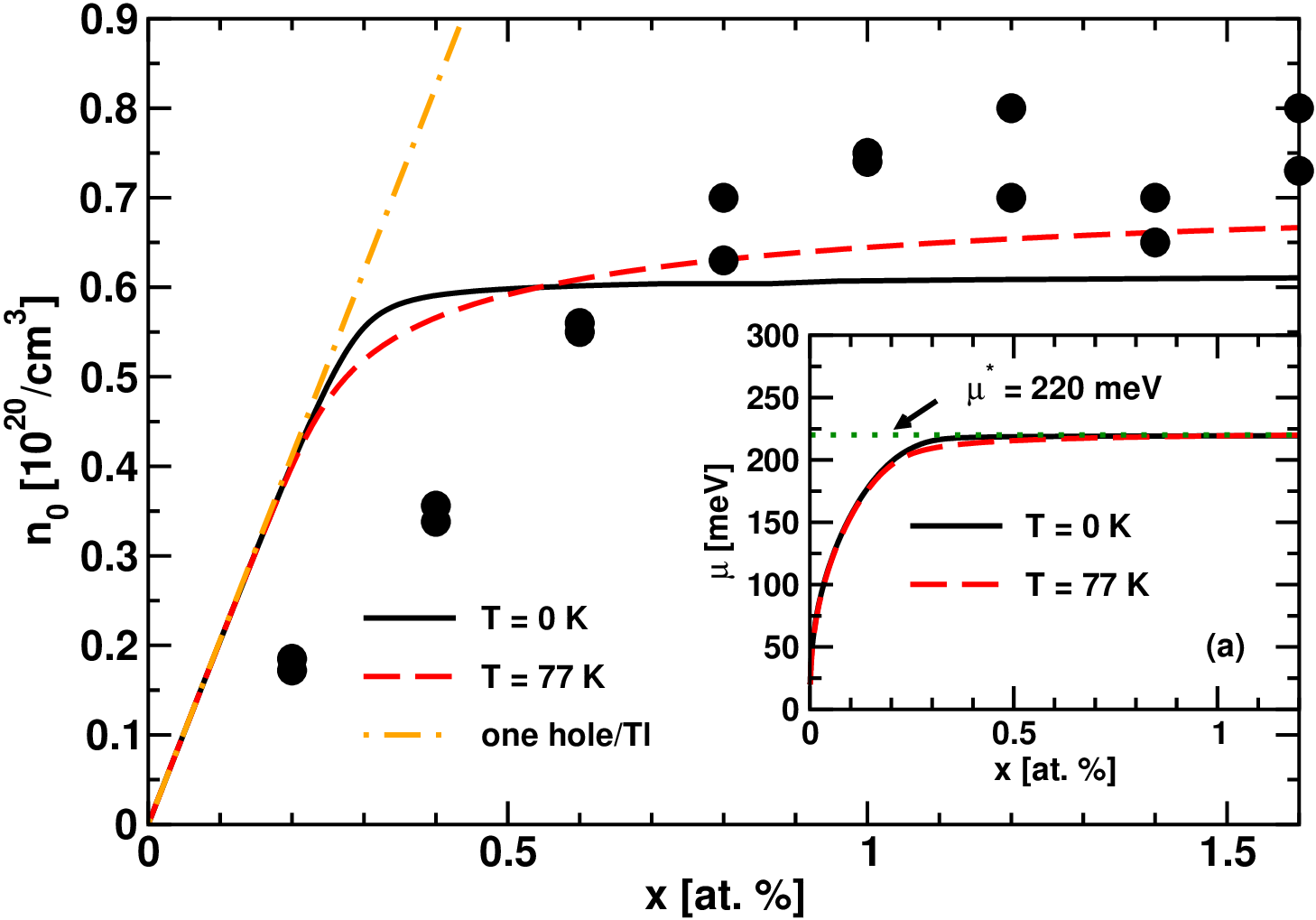}
\caption{Hole carrier density $n_{0}$ versus {\rm Tl} doping $x$ in atomic \% (at. \%) for 
$T=0\, {\rm K}$ and $T=77\, {\rm K}$. The dot-dashed line is the hypothetical $n_{0}$ 
for one hole per {\rm Tl}. Filled circles are experimental data from Hall number 
measurements at $T=77\,{\rm K}$ \cite{murakami.96}.
Inset (a): hole chemical potential $\mu$ versus $x$ at $T=0\, {\rm K}$ 
and $T=77\, {\rm K}$ (dotted line indicates value of the pinned chemical potential 
$\mu^{*}=220\,{\rm meV}$. The value of the charge splitting $\delta(\mu)=2(\mu-\mu^{*})$
can also be read off from this plot.
\label{fig1}}
\end{figure}
Figure~\ref{fig1} shows the hole carrier density $n_{0}(x)$ versus Tl concentration $x$ (in \%) 
at $T=0$ and at $T=77\,{\rm K}$.
For $x<x_{c}\approx 0.3$, $n_{0}$ is linear in $x$, i.e. each {\rm Tl} contributes 
one hole, as in the case of vacancies on {\rm Pb} sites (dot-dashed curve in Fig.~\ref{fig1}). 
For $x> x_{c} \approx 0.3$, $n_{0}(x)$ saturates rapidly with increasing $x$ 
for $T=0$ and more slowly at finite temperature. The hole chemical
potential $\mu$, shown in Fig.~\ref{fig1}a, grows non-linearly with $x$ at $x<x_{c}$ and
rapidly approaches the value $\mu^{*}=200\,{\rm meV}$ for $x>x_{c}$, both at $T=0$ and at $T=77\,{\rm K}$.
Both of these results are in good qualitative agreement with experimental
data \cite{murakami.96} (see \cite{costi.12} for comparisons to more data \cite{matsushita.06}).  
Since we use the realistic band structure, we are also able to obtain quantitative agreement for
the saturation density $n_{0}\approx 0.7 \times 10^{20}/{\rm cm}^{3}$. The self-compensation
effect at $x\gg x_{c}$ is a ``smoking gun'' signature for the charge Kondo state: on entering this
state, the Tl ions fluctuate between Tl$^{1+}$ and Tl$^{3+}$ so the average valence Tl$^{2+}$ is
neither a donor nor an acceptor and the carrier density ceases to increase for $x>x_{c}$. Counterdoping
with donor ions, removes holes, increases $\mu$ and the charge splitting and consequently
destroys the charge Kondo state. Indeed, counterdoping with In, a donor ion, has been shown
to destroy the charge Kondo anomalies in the resistivity and the onset of superconductivity 
\cite{erickson.10}

\section{Tl $s$-Electron Spectral Function}
\label{sec:7}
\begin{figure}
\includegraphics[width=\linewidth,clip]{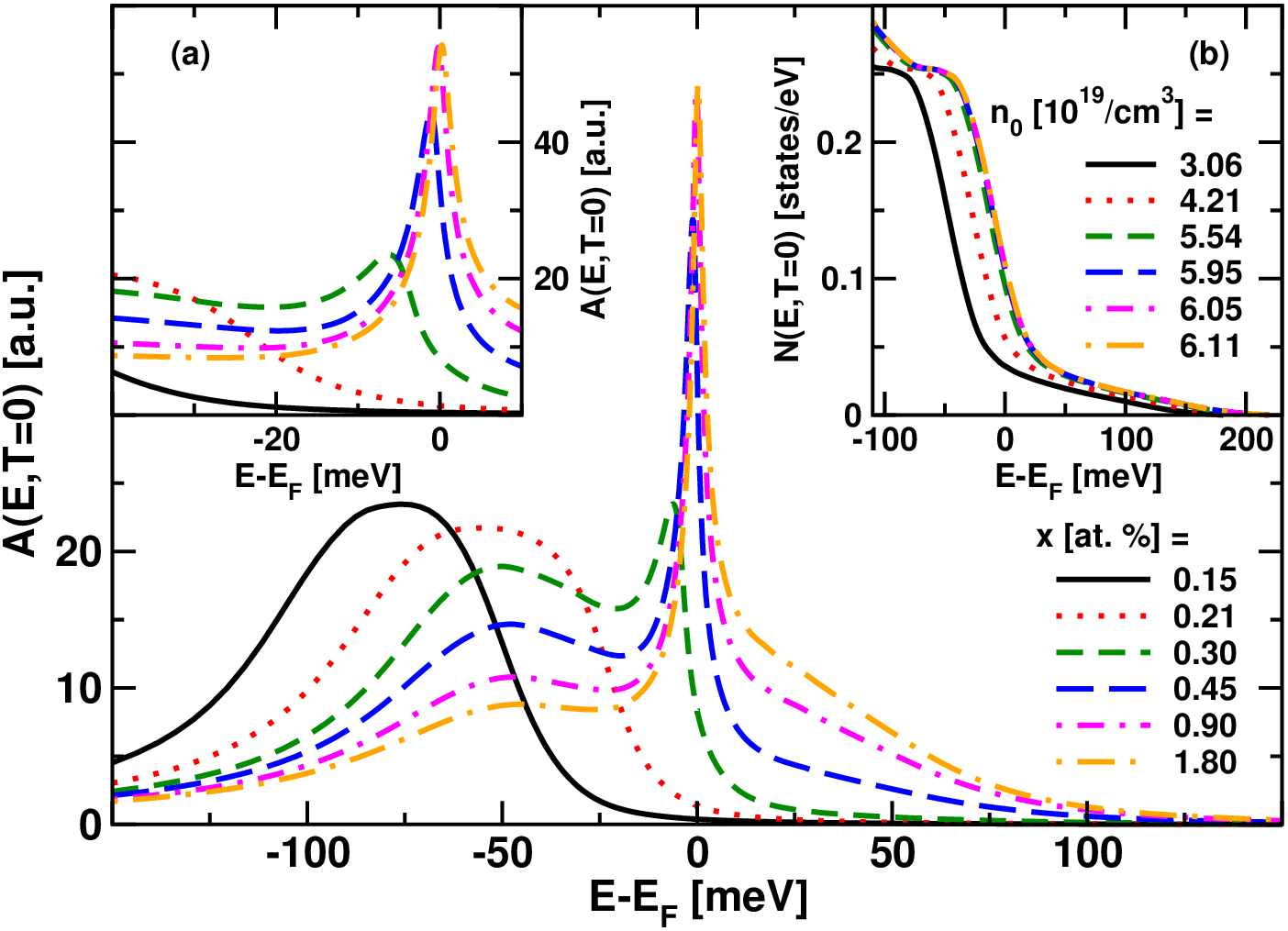}
\caption{Main panel: $T=0$ spectral function $A(E,T=0)$ versus energy 
$E-E_{\rm F}$ for a range of {\rm Tl} dopings $x$. Here, $E_{\rm F}=E_{\rm F}(x)$ is the self-consistently
calculated Fermi level for each $x$. Inset (a): region near $E=E_{\rm F}$ showing the
charge Kondo resonance. Inset (b): valence band local density of states ${\cal N}(E)$ 
versus $E-E_{\rm F}$ for $x$ as in the main panel. The conduction band (not shown) lies 
190\,{\rm meV} above the top of the valence band and is not important for the present
study. The hole carrier densities $n_{0}$ for each $x$, obtained
from Fig.~\ref{fig1}, are given in the legend.
\label{fig2}}
\end{figure}
Figure~\ref{fig2} shows the evolution with {\rm Tl} doping of the single-particle spectral function
$A(E,T=0)$ at zero temperature. For small doping, the hole chemical potential $\mu$ 
lies above $\mu^{*}$, within the shallow part of the valence band density of states 
(see inset Fig.~\ref{fig2}b), and consequently the splitting $\delta(\mu)=2(\mu-\mu^{*})$
is large. A large charge splitting in the negative-$U$ Anderson model acts like a large Zeeman
splitting in the corresponding positive-$U$ model. Consequently for $x\ll x_{c}$, 
the spectral function is strongly polarized, with most weight lying in a Hubbard satellite 
peak far below the Fermi level $E_{\rm F}$ with no other peaks in the spectral function. 
With increasing $x$, the charge splitting $\delta(\mu)$ decreases, the spectral function
becomes less polarized, and for $x\ge x_{c}$, in addition to the Hubbard peak discussed above,
a charge Kondo resonance develops close to $E_{\rm F}$. The latter is a result of dynamic 
valence fluctuations between the almost
degenerate Tl$^{1+}$ and Tl$^{3+}$ configurations. Point contact measurements for PbTe samples
doped with $x>x_{c}$  show the existence of two quasi-localized states \cite{murakami.96}, a narrow
one of width $6\,{\rm meV}$ close to the Fermi level and a broader one of width $12\, {\rm meV}$ further
below the Fermi level, in broad agreement with our theoretical calculations \cite{costi.12}. Indeed,
in \cite{costi.12}, this interpretation of the experiments was used to estimate $U$ from
the separation of the lower Hubbard band from the Kondo resonance.

\section{Resistivity Anomaly}
\label{sec:8}
\begin{figure}
\includegraphics[width=\linewidth,clip]{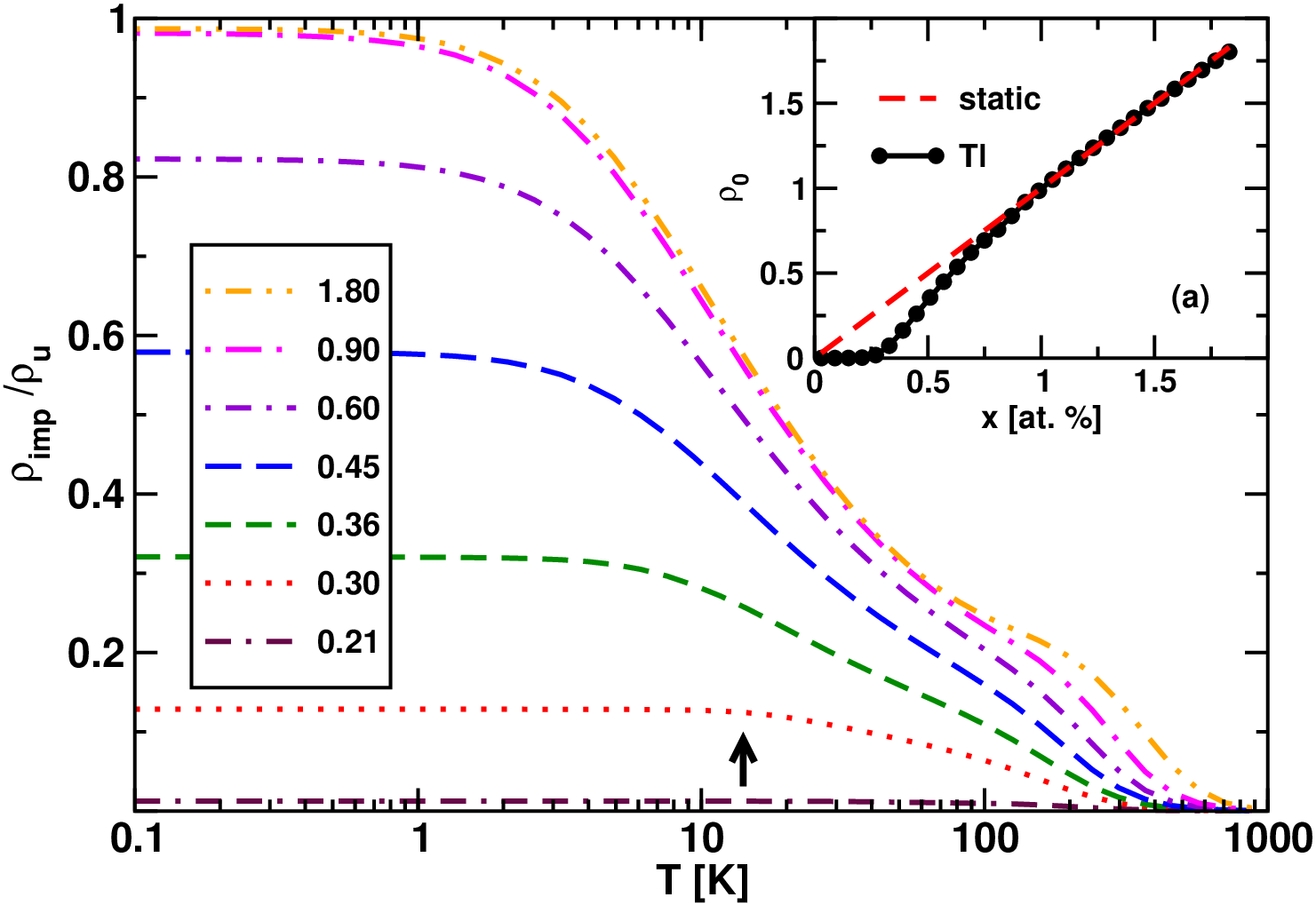}
\caption{Normalized impurity contribution to the 
resistivity $\rho_{\rm imp}(T)/\rho_{u}$ versus temperature $T$ and
a range of {\rm Tl} concentrations $x$ (in \%, see legend). Note that
for $T>10\, {\rm K}$, the total resistivity will be dominated by the
phonon contribution, not shown in this figure. 
$\rho_{u}=2c_{\rm imp}/(e^{2}\pi \hbar N_{F}^{2}\sigma_{0})$ 
is the residual resistivity for unitary scatterers
with $N_{F}=N(\mu^{*})$ the density of valence band states at $\mu=\mu^{*}$ and
$\sigma_{0}$ a material dependent constant. 
The vertical arrow indicates the Kondo temperature $T_{\rm K}=14\,{\rm K}$ at the charge Kondo
degeneracy point. Inset (a): residual resistivity 
$\rho_{0}=x\rho_{\rm imp}(T=0)/\rho_{u}$ versus $x$ (filled circles). 
Dashed-line: residual resistivity for static impurities in place of {\rm Tl}.
\label{fig3}}
\end{figure}
The temperature and doping dependence of the impurity resistivity, $\rho_{\rm imp}$,
is shown in Fig.~\ref{fig3}. 
For $x<x_{c}$, Tl impurities act as acceptors with a well-defined valence
state (Tl$^{1+}$). They therefore act as weak potential scatterers 
and consequently the resistivity is much below the unitary value, as
seen in Fig.~\ref{fig3}. For $x>x_{c}$,
dynamic fluctuations between the nearly degenerate Tl$^{1+}$ and Tl$^{3+}$ states
leads to the charge Kondo effect and $\rho_{\rm imp}$ approaches the resistivity 
for unitary scatterers at $T=0$. For $x>x_{c}$, $\rho_{\rm imp}$ is well described by the 
{\em spin} Kondo resistivity \cite{costi.12} 
with a logarithmic form around $T\approx T_{\rm K}^{\rm eff}$, where $T_{\rm K}^{\rm eff}$ is 
an effective Kondo scale, and $T^{2}$ Fermi liquid corrections 
at low $T\ll T_{\rm K}^{\rm eff}$, in qualitative
agreement with experiment \cite{matsushita.05}. The effective 
Kondo scale $T_{\rm K}^{\rm eff}$ is a function of the charge splitting 
$\delta(\mu)$ and $T_{\rm K}$, and approaches the true Kondo scale $T_{\rm K}$
only asymptotically for $x\gg x_{c}$ (see legend to Fig.~\ref{fig3}). 
Finally, Fig.~\ref{fig3}a shows that the impurity residual resistivity is
significant only when the charge Kondo effect is operative, i.e. for $x>x_{c}$, in
qualitative agreement with experiment \cite{matsushita.05,matusiak.09}.

\section{Thermopower}
\label{sec:9}
\begin{figure}
\includegraphics[width=\linewidth,clip]{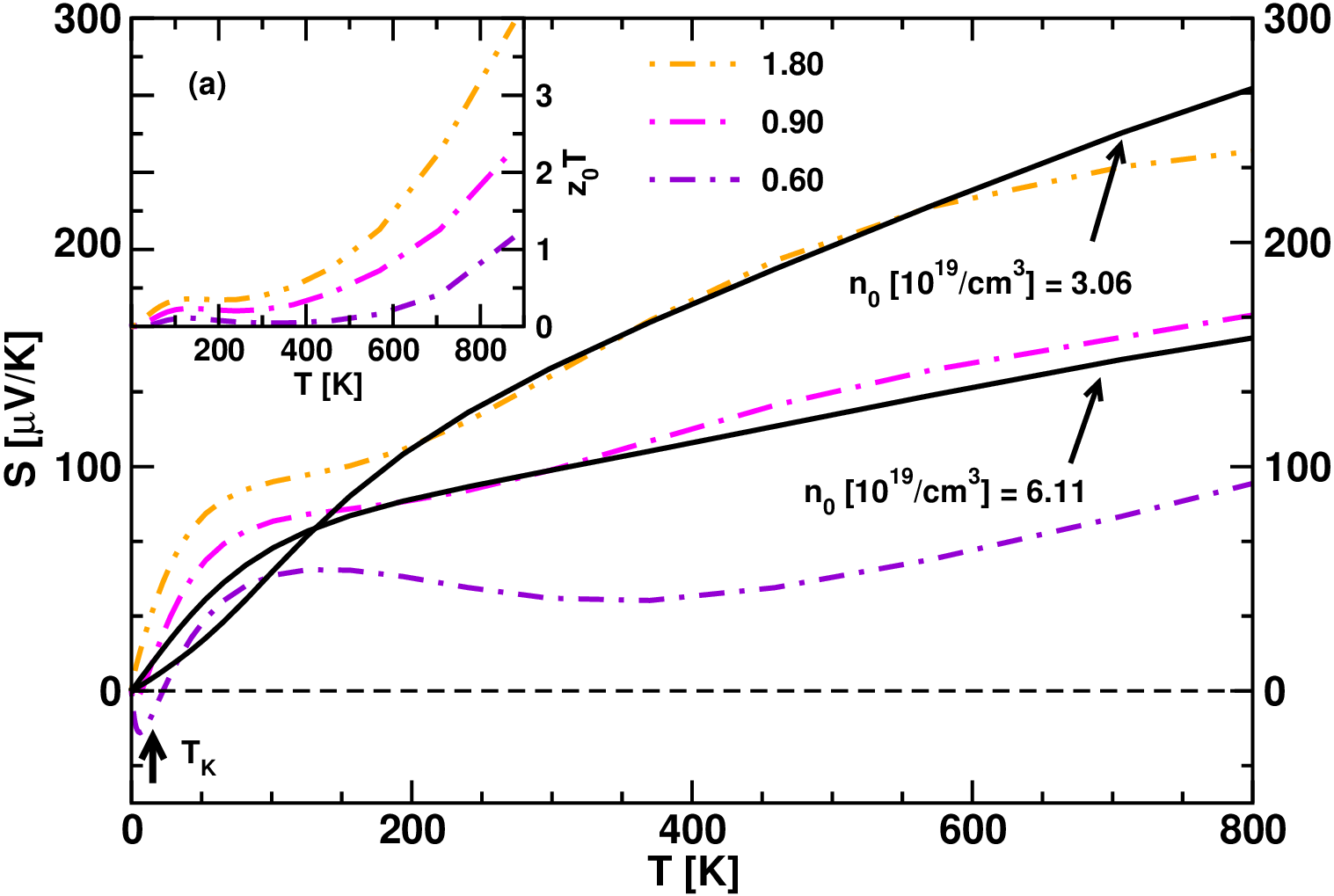}
\caption{Main panel: Thermopower $S$ versus temperature at different
{\rm Tl} concentrations $x$ (in \%, see legend) in the highly doped regime $x>x_{c}$. 
The two solid lines labeled by $n_{0}$ correspond to $S$ for the two highest 
doped cases (i.e. for $x=0.9$ and $x=1.8$) assuming a constant transport time.  
Inset (a) shows the temperature dependence of the dimensionless electronic 
figure of merit $z_{0}T$ for $x>x_{c}=0.3$ (in \%, see legends). 
This is indicative of trends only, since the lattice contribution to the thermal conductivity
has been neglected in $z_{0}T$. The vertical arrow denotes the particular choice of
Kondo temperature $T_{\rm K}=14\,{\rm K}$ made in this paper (see discussion in
Sec.~\ref{sec:4}).
\label{fig4}
}
\end{figure}

In Figure~\ref{fig4} we show the temperature dependence of the thermopower $S(T)$ 
for several Tl concentrations in the highly doped regime. At low temperature $T\ll T_{\rm K}$,
the thermopower can be obtained from a Sommerfeld expansion of Eq.~(\ref{eq: thermopower}),
\begin{equation}
\label{eq:sommerfeld}
S(T)=-\frac{\pi^{2}k_{B}}{3|e|}k_{B}T
\left(\frac{{\cal N}'(E_{\rm F})}{{\cal N}(E_{\rm F})} 
-\frac{A'(E_{\rm F})}{A(E_{\rm F})}\right),
\end{equation}
where the first term is the band contribution to the thermopower 
and the second term involving the Tl $s$-electron spectral function is that due
to the charge Kondo effect of the Tl impurities. 
Since Tl ions act as acceptors, the Tl $s$-electron spectral function, $A(E)$, has most of
its weight below $E_{\rm F}$ and its slope at $E_{\rm F}$, like that of ${\cal N}(E)$, is negative
(see Fig.~\ref{fig2} and Fig.~\ref{fig2}b). Consequently, at low temperature, 
the charge Kondo and band contributions to the thermopower are both $p$-type and 
compete with each other. This implies that thermopower can undergo a sign change at
low temperature, depending on details, such as the Tl doping level. Recent measurements 
show the occurrence of such sign changes at low Tl dopings \cite{matusiak.09} consistent
with a charge Kondo effect. However, our effective single band Anderson model for Tl ions
in PbTe is not accurate enough to describe the details of these sign changes, for the following reason. 
The valence $p$-band of PbTe actually consists of two sub-bands, light and heavy hole 
bands due to different hole pockets in  the Brillouin zone \cite{sitter.77}. These
bands with densities of states ${\cal N}_{1}(E)$ and ${\cal N}_{2}(E)$ 
become successively occupied with increasing Tl doping, and for $x\approx x_{c}$
there is significant population of both. For simplicity, in Sec.~\ref{sec:2} the effect
of both bands was approximated only in the density of states 
${\cal N}(E)={\cal N}_{1}(E)+{\cal N}_{2}(E)$.  For transport
properties, such as thermopower, which depend sensitively on the relative populations and  
mobilities of such sub-bands, an effective one band model is, in general, not 
adequate. We expect, however, that it is approximately correct in the high doping 
limit when transport is mainly due carriers in the heavy hole band. 
In this limit, and at temperatures $T\gg T_{\rm K}$, when the charge Kondo effect is suppressed,
our results for $S$ should reproduce those from just the band contribution (${\cal N}(E)$) 
within a constant relaxation time approximation $\tau(E,T)=\tau_{0}$ \cite{singh.10}.  
Fig.~\ref{fig4} shows that this is indeed the case for $x>0.9\%$ and $T\ge 175 {\rm K}$.
The large thermopower in the range $300-800 {\rm K}$ increases with decreasing
carrier density, reflecting the enhancement of ${\cal N}'(E_{\rm F})/{\cal N}(E_{\rm F})$ with
increasing $E_{\rm F}$ (i.e. {\em decreasing} hole chemical potential $\mu$). 
In Fig.~\ref{fig4}b we show the electronic figure of merit
$z_{0}T = PT/\kappa_{e}$, where $\kappa_{e}$ is the electronic thermal conductivity. The
quantity $z_{0}T$ is an upper bound to the true figure of merit $ZT$. 
We find $z_{0}T=2$ and $z_{0}T=4$ at 
$T=800 {\rm K}$ for $x\approx 1\%$ and $x \approx 2\%$, respectively, 
approximately twice larger than the reported measured values of $ZT$ at the same 
temperature and doping levels\cite{heremans.08}. This is reasonable, given our neglect
of the lattice thermal conductivity.

\begin{figure}
\includegraphics[width=\linewidth,clip]{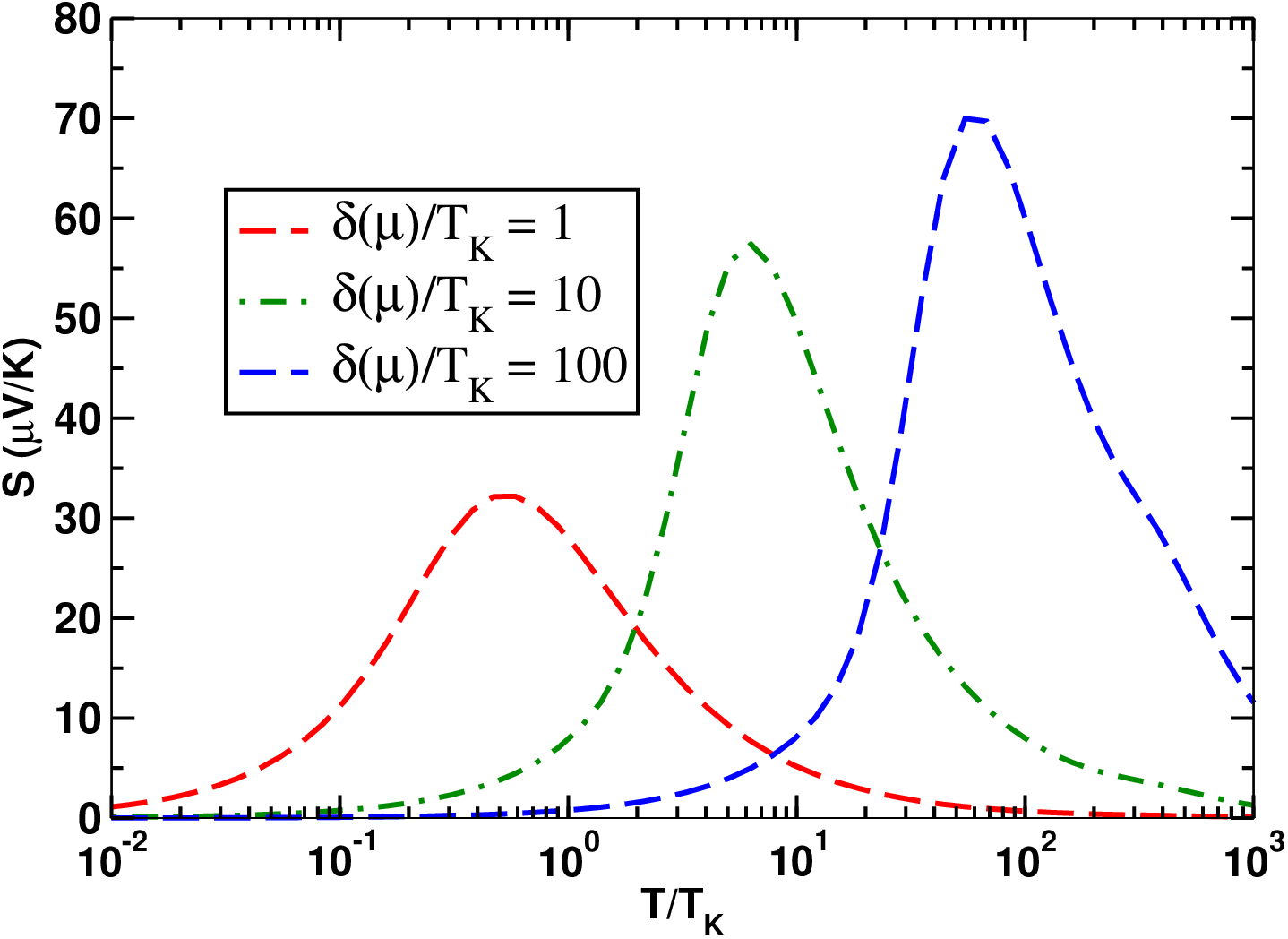}
\caption{Thermopower $S$ versus reduced temperature $T/T_{\rm K}$ 
of the negative-$U$ Anderson impurity model
in a metallic host \cite{andergassen.11} for $|U|/\Delta_{0}=8$
and several values of the charge splitting $\delta(\mu)$. The latter is
defined as the splitting between the empty and doubly occupied states of
the  negative-$U$ Anderson impurity model and is analogous to the 
self-consistently calculated $\delta(\mu)$ for the Tl impurities in 
PbTe discussed in this paper. $T_{\rm K}$ denotes the Kondo scale.
The charge Kondo effect provides a mechanism for large thermopowers
of order $k_{\rm B}/e\approx 76\mu {\rm V/K}$ at charge splittings 
$\delta(\mu)/T_{\rm K}\ge 1$. Since $T_{\rm K}$ can be made small, due to
its exponential dependence on $U$ and $\Delta_{0}$, such
splittings are accessible in potential realizations of the charge Kondo effect,
e.g. in negative-$U$ centers in semiconductors 
or in molecular junctions \cite{andergassen.11}. 
\label{fig5}
}
\end{figure}

Within our model calculations, the charge Kondo contribution to the 
thermopower at low temperatures competes with the main contribution 
coming from the band instead of supplementing it. However, the 
contribution from the charge Kondo effect is generically quite large, 
and can reach values of order $76{\mu {\rm V/K}}$ for 
splittings $\delta(\mu)/T_{\rm K} > 1$. This is indicated in Fig.~\ref{fig5}
for the charge Kondo thermopower of the negative-$U$ Anderson impurity model 
\cite{andergassen.11} using a flat band appropriate for impurities in 
metallic systems. Hence, in principle, impurities exhibiting the charge Kondo 
effect could enhance the overall thermopower by the above value. This could
offer a route to further improving the thermoelectric properties of appropriate
materials.

\section{Conclusions}
\label{sec:10}
In summary, we investigated the low temperature  properties of 
Pb$_{1-x}$Tl$_{x}$Te within a model of Tl impurities acting as negative $U$ centers. 
Our NRG calculations explain a number of low
temperature anomalies of Pb$_{1-x}$Tl$_{x}$Te, including the qualitatively 
different behavior below and above the critical concentration $x_{c}$, 
where $x_{c}\approx 0.3\,\%$.  They support the suggestion that the charge Kondo effect is realized in 
Pb$_{1-x}$Tl$_{x}$Te \cite{matsushita.05,dzero.05}. At $x=x_{c}$, two 
nonmagnetic valence states of Tl become almost degenerate and the ensuing pseudospin 
charge Kondo effect results in a Kondo anomaly in the resistivity for $x>x_{c}$ and a 
residual resistivity approximately linear in $x$.
Our results for these quantities and the carrier density $n_{0}(x)$
are in good qualitative agreement with experiments 
\cite{matsushita.05,matsushita.06,matusiak.09,murakami.96}. 
For the Tl $s$-electron spectral function, we predict that one peak should be 
present far below $E_{\rm F}$ for $x<x_{c}$ and that a second temperature dependent 
Kondo resonance peak develops close to, but below $E_{\rm F}$, 
on increasing $x$ above $x_{c}$. This provides a new interpretation of measured 
tunneling spectra\cite{murakami.96}, which could be tested by temperature 
dependent studies of tunneling or photoemission spectra. We also showed that
the competing charge Kondo and band contributions to the low temperature thermopower 
imply that there are sign changes in the thermopower at low Tl concentrations. Investigating
these in detail, particularly at $x<0.3\%$, could shed further light on the charge Kondo 
effect in this system. In the future, it would
be interesting to extend this work to investigate the effects of disorder, phonons, light and
heavy hole bands, and non-resonant scattering channels on the thermoelectric
properties of Tl doped PbTe.

\begin{acknowledgement}
We thank K. M. Seemann, D. J. Singh, H. Murakami, P. Coleman, 
G. Kotliar, J. Schmalian, and I. R. Fisher for discussions and 
D. J. Singh, H. Murakami and I. R. Fisher for data \cite{singh.10,matsushita.06}. 
V.Z. acknowledges support by Croatian MZOS Grant No.0035-0352843-2849, 
NSF Grant DMR-1006605 and Forschungszentrum J\"{u}lich.
T. A. C. acknowledges supercomputer support from the John von Neumann Institute for 
Computing (J\"{u}lich).
\end{acknowledgement}

%
%

\end{document}